\documentclass{aa}
\usepackage{psfig}

\def\la{\;
\raise0.3ex\hbox{$<$\kern-0.75em\raise-1.1ex\hbox{$\sim$}}\; }
\def\ga{\;
\raise0.3ex\hbox{$>$\kern-0.75em\raise-1.1ex\hbox{$\sim$}}\; }

\newcommand{\zabs}{$z_{\rm abs}\,$}

\newcommand{\kms}{km~s$^{-1}\,$}
\newcommand{\ms}{m~s$^{-1}\,$}
\newcommand{\cm}{cm$^{-2}\,$}

\newcommand{\iso}{$^{12}$C/$^{13}$C$\;$}

\begin{document}

\title{VLT/UVES constraints on the carbon isotope ratio \iso\\
at $z=1.15$ toward the quasar \object{HE 0515--4414}\thanks{Based on 
observations performed at the VLT Kueyen telescope (ESO, Paranal, Chile). 
The data are retrieved  from the ESO/ST-ECF Science Archive Facility.
}
}
\author{
S.~A.~Levshakov\inst{1}
\and
M.~Centuri\'on\inst{2}
\and
P.~Molaro\inst{2,3}
\and
M.~V.~Kostina\inst{4}
}
\offprints{S.~A.~Levshakov
\protect \\lev@astro.ioffe.rssi.ru}
\institute{
Department of Theoretical Astrophysics,
Ioffe Physico-Technical Institute, 194021 St.~Petersburg, Russia
\and
Osservatorio Astronomico di Trieste, Via G. B. Tiepolo 11,
34131 Trieste, Italy
\and
Observatoire de Paris 61, avenue de l'Observatoire, 75014 Paris, France
\and
Sobolev Astronomical Institute, St.~Petersburg State University,
198504 St.~Petersburg, Russia
}

\date{Received 00  / Accepted 00 }

\abstract{}
{We analyzed the \ion{C}{i} lines associated with the damped
Ly$\alpha$ system observed at \zabs = 1.15 in the spectrum 
of \object{HE 0515--4414} to derive the \iso ratio.}
{ The spectrum was obtained by means of the UV-Visual Echelle
Spectrograph (UVES) at the ESO Very Large Telescope (VLT). }
{The obtained lower limit \iso~$> 80$ (2$\sigma$ C.L.) shows 
for the first time that the abundance of $^{13}$C in the extragalactic
intervening clouds is very low. 
This rules out a significant contribution from 
intermediate-mass stars to the chemical evolution of matter
sampled by this line of sight.
The estimated low amount of $^{13}$C is in agreement with
low abundances of nitrogen observed in damped Ly$\alpha$ systems~--
the element produced in the same nuclear cycles and from about the same
stars as $^{13}$C.}
{}

 
\authorrunning{S. A. Levshakov et al.}
\titlerunning{\iso\ at $z=1.15$ toward \object{HE 0515--4414}}
\maketitle

\section{Introduction}

The evolution of the chemical composition of matter in the Universe
is closely related to the history of star formation and destruction,
stellar nucleosynthesis and enrichment of the interstellar/intergalactic 
medium (ISM/IGM) with processed material.
The behavior of the stable CNO isotopes during the cosmic time is 
of particular interest  since their ratios trace the production of 
{\it primary} and {\it secondary} elements\footnote{A chemical element 
is called primary/secondary if its mass fraction ejected from a star
is insensitive/ sensitive to the original metallicity of the
star (Talbot \& Arnett 1974).}
which depends on the principal chemical evolution parameters: 
the stellar initial mass function (IMF), 
the rate of the mass loss from
evolved stars, and the dredge-up episodes leading to the
mixture of the surface layers with deeper layers of the star
(Wannier 1980; Renzini \& Voli 1981;  
Marigo 2001; Meynet et al. 2005).

$^{12}$C is a primary product of stellar nucleosynthesis and
is formed in the triple-$\alpha$
process during hydrostatic helium burning  
(van den Hoek \& Groenewegen 1997; El Eid 2005).
$^{13}$C is supposed to have mainly a secondary
origin, and is produced in the hydrogen burning shell when the CN cycle
converts $^{12}$C into $^{13}$C (Wannier 1980). 
However, chemical evolution models considered by Prantzos et al. (1996)
match observations better if a mixture of primary+secondary
origin in intermediate-mass stars is assumed for $^{13}$C.
Evolution of massive rotating stars at very low metallicities
may also contribute to primary $^{13}$C (Meynet et al. 2005).
The predicted values for the ratio ${\cal R}$~=\iso in the rotationally
enhanced winds diluted with the supernova ejects are between 100 and 4000, 
whereas the $^{13}$C yields of massive non-rotating stars are negligible,
${\cal R} \sim 3\times10^8$. Besides, Meynet et al. show that
the rotating AGB and massive stars have 
about the same effects on the isotope production.
The difference between them is
only in the isotope composition of the massive star wind material
and the AGB star envelopes: the former is characterized by very
low values of ${\cal R} \sim 3-5$, while the latter have 
${\cal R}$ ranging between 19 and 2500 (the lower ${\cal R}$ values 
correspond to the most massive AGB stars).

Chemical evolution models predict a decrease of the
isotope ratio \iso with time and an increase 
with galactocentric distance at a fixed time (Audouze et al. 1975;
Dearborn et al. 1978; Tosi 1982; Romano \& Matteucci 2003). 

For instance, a photospheric solar ratio ${\cal R} = 95\pm5$ 
(Asplund et al. 2005),
representative of the local ISM 5 billion years ago, is higher than the
present value ${\cal R} \sim 60-70$ obtained through optical, UV, and 
IR absorption line observations as well as radio emission line measurements 
(see, e.g., Hawkins \& Jura 1987;
Centuri\'on \& Vladilo 1991; Centuri\'on et al. 1995;
Goto et al. 2003; and references therein). 
It should be noted, 
however, that on scales of $\sim 100$~pc 
the local ISM is probably chemically inhomogeneous
(Casassus et al. 2005) making this difference uncertain.

The high resolution observations ($\lambda/\delta\lambda \ga 50\,000$)
of quasar absorption-line spectra available nowadays 
at large telescopes allow us to probe
the isotope composition of the intervening damped Ly$\alpha$ (DLA) systems 
through the analysis of \ion{C}{i} lines, and, hence, to perform an
important test of models of stellar nucleosynthesis outside the Milky Way. 
The most convenient \ion{C}{i} transitions
are from the $2u$ ($\lambda1657$ \AA) and $3u$ ($\lambda1561$ \AA) 
multiplets (Morton 2003) which cover the redshift
range from $z \sim 1$ to $z \sim 3.5$ 
(look-back time $\sim 7.7-11.8$ Gyr)\footnote{Assuming 
the cosmological expansion parameterized by
$t_0 = 13.8$~Gyr, $h = 0.7$, $\Omega_m = 0.3$, and $\Omega_\Lambda = 0.7$.} 
in optical spectra of distant QSOs.

The analysis of the isotopic composition of the DLA systems
is also connected to the interpretation 
of the hypothetical variation of the fine-structure constant
($\alpha\,\, \equiv e^2/\hbar c$) at early cosmological epochs 
(Ashenfelter et al. 2004a,b; Fenner et al. 2005,  hereafter FMG).

In the present letter we report constraints on the carbon isotope abundance 
in the \zabs = 1.15 sub-DLA system toward 
\object{HE 0515--4414} (Reimers et al. 1998).
The metallicity of this sub-DLA is found to be lower than solar:
[Zn/H] $\simeq -0.99$\footnote{[A/B] $\equiv \log(N_{\rm A}/N_{\rm B}) -
\log(N_{\rm A}/N_{\rm B})_\odot$.} 
according to de la Varga et al. (2000), or
[Zn/H] $\simeq -0.49$ as measured by Reimers et al. (2003) who revised the
\ion{H}{i} column density.
Sects.~2 and 3 describe the observations and the analysis.
The obtained results and their astrophysical implications 
are discussed in Sect.~4.

\begin{figure}[t]
\vspace{0.0cm}
\hspace{-2.0cm}\psfig{figure=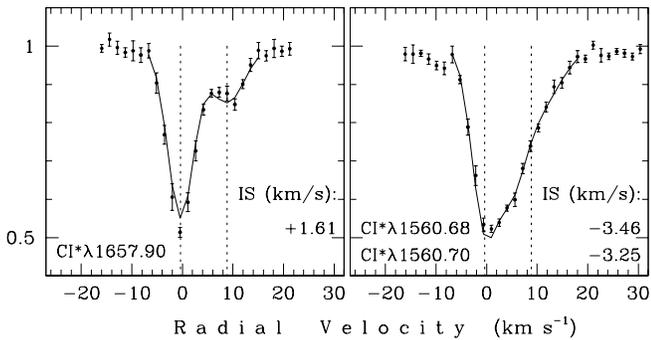,height=13.0cm,width=13.0cm}
\vspace{-8.6cm}
\caption[]{
Normalized intensities (dots with $1\sigma$ error bars) of the
\ion{C}{i} lines selected from the VLT/UVES
spectrum of \object{HE 0515--4414}. 
The zero radial velocity is fixed at $z = 1.150789$ (QBR).
The over-plotted smooth curves show the best
fitted synthetic profiles ($\chi^2_{\rm min} = 1.250$, the number
of degrees of freedom $\nu = 26$). 
The dashed vertical lines mark positions of the \ion{C}{i}
sub-components. Note that the $\lambda1560$ absorption feature consists of
two sub-components with different oscillator strengths:
$f_{1560.68} = 0.0581$ and $f_{1560.70} = 0.0193$.
The $^{12}$\ion{C}{i}$^\ast$ oscillator strengths, 
laboratory vacuum wavelengths and isotopic shifts (IS) are taken
from Morton (2003)
}
\label{fig1}
\end{figure}

\section{Observations and data reduction}

The observations were acquired with the UV-Visual Echelle Spectrograph (UVES) 
at the VLT 8.2~m telescope at Paranal, Chile, and the spectral data were
retrieved from the ESO archive. 
The seven selected exposures were described in detail in
Levshakov et al. (2005a, hereafter Paper~I).

The \ion{C}{i} lines analyzed in the present paper are located near 
the center of the echelle orders. 
This minimizes possible distortions of the line profiles
caused by the decreasing spectral sensitivity at the edges 
of echelle orders.

Following Paper~I, where all details of the data reduction are given, 
we work with the reduced spectra with the 
original pixel sizes in wavelength. 
The residuals of the calibrations give $\sigma_{\rm rms} \la 1$ m\AA.
The observed wavelength scale of each spectrum was transformed into vacuum,
heliocentric wavelength scale (Edl\'en 1966). 
The instrumental profile is dominated in this case by the slit width 
($0.8$ arcsec). 
The spectral resolution
calculated from the narrow lines of the arc spectra 
is FWHM~$= 5.60\pm0.10$ \kms. 

After the normalization to the local continuum,
the spectra from different exposures were rebinned with the step
equal to the mean pixel size. To eliminate instrumental velocity
calibration errors, we used the first exposure as a reference frame
and calculated residual differences in the radial velocity offsets of other
exposures through the cross-correlation analysis. 
The aligned \ion{C}{i} spectra were
co-added with weights inversely proportional to $\sigma^2_{\rm cont}$.   
The resulting average \ion{C}{i} profiles are shown in Fig.~1 by dots with 
1$\sigma$ error bars.

\section{Analysis}

The key points in the study of the isotopic abundance from 
the UV \ion{C}{i} lines are as follows: 
 
\smallskip\noindent
1.\, At high redshift, \ion{C}{i}
is observed in the most dense and cool
sub-components of DLA systems where H$_2$ lines are usually
also detected. The kinetic temperature in such \ion{C}{i}-bearing 
clouds is low. For our particular case $T_{\rm kin} \sim 200$~K
(Reimers et al. 2003). It means that the thermal width of the 
carbon lines is small, $b_{\rm th} \sim 0.5$ \kms, and the line profiles
can be very narrow if the turbulent broadening is not significant.

\smallskip\noindent
2.\, The isotopic shift (IS)\footnote{Defined as IS~= 
$c\,(\lambda_{13} - \lambda_{12})/\lambda_{12}$ (in \kms).}
between $^{12}$C and $^{13}$C lines
shows different values and signs (Morton 2003).
For instance, IS$_{1657.907/1657.916} = 1.61$ \kms, whereas
IS$_{1560.682/1560.664} = -3.46$ \kms.
Although with the spectral resolution of FWHM = 5.6 \kms these lines 
cannot be completely resolved, the presence of $^{13}$C can shift and
broaden $^{12}$C lines leading to the disparity of line
positions from different multiplets.
The effect is small when \ion{C}{i} lines are optically thin. 
The centroid of a pair of lines $^{12}$C+$^{13}$C  is shifted
by $\Delta v = {\rm IS}/(1+{\cal R})$ with respect to the position of the
$^{12}$C line.
Suggesting ${\cal R} = 60$, this gives for the 1657.907 and
1560.682 lines $\Delta v = +0.026$ and  $-0.057$ \kms,
making the disparity of line positions compared to the assumption of pure
$^{12}$C equal to 0.083 \kms. However, the disparity increases with increasing
optical depth. 
For saturated lines with $\tau_0$($^{13}$C)~$> 1$  
and a pure rectangular velocity profile, the short wavelength 
side of this box-shaped absorption can be displaced by +IS or --IS.
The centroid of $^{12}$C+$^{13}$C is then shifted by
half of IS, irrespective of ${\cal R}$. In this case the
disparity is equal to 2.535 \kms, which already 
detectable at our resolution.
Note that for \object{HE 0515-4414}
the line centering can be done with an accuracy 
better than 1/10 pixel size which is about 150 \ms (Paper~I).
Since the strengths of carbon lines from the UV multiplets cover
a wide dynamical range, the saturation effects can be correctly taken
into account.

\smallskip\noindent
3.\, Since the atomic carbon isotopic abundance seems to be
unaffected by depletion, 
the measured ratio \iso reflects the carbon isotopic ratio at
a given redshift.

\begin{figure}[t]
\vspace{-1.0cm}
\hspace{-1.0cm}\psfig{figure=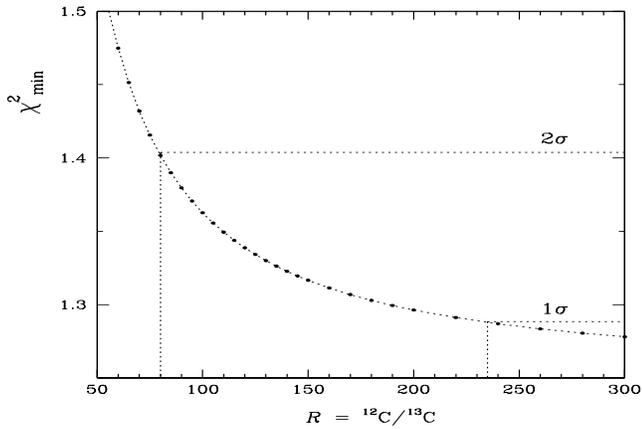,height=8.0cm,width=11.5cm}
\vspace{-1.6cm}
\caption[]{
Confidence intervals in the ``$\chi^2$-${\cal R}$'' plane
calculated from the simultaneous fit of the
\ion{C}{i}$^\ast$ lines shown in Fig.~1. 
}
\label{fig2}
\end{figure}

\smallskip
The VLT/UVES spectrum of \object{HE 0515--4414}
reveals multi-component \ion{C}{i} profiles at \zabs = 1.15 (Fig.~1).
Their diagnostic was performed by 
de la Varga et al. (2000) and later by Quast et al. (2002, 
hereafter QBR).
All \ion{C}{i} lines are well described by 
a two-component Voigt profile model
(see Fig.~1 in QBR). Besides, QBR found  
that the sub-components of the \ion{C}{i}
lines arising from the $J=0, 1$, 
and 2 levels of the ground state have different
relative strengths indicating inhomogeneous physical conditions in the
\ion{C}{i}-bearing cloud.

The heterogeneity of the absorbing medium may cause small relative
velocity shifts between \ion{C}{i} transitions 
with different $J$ similar to the shifts 
observed between H$_2$ lines in the direction of $\zeta$~Ori~A (Jenkins \& 
Peimbert 1977). This would add an additional noise in the measurements
of the disparity of the \ion{C}{i} line positions.
To avoid this putative effect we selected from the observed eight \ion{C}{i}
lines (QBR) only two $J=1$ transitions which show most pronounced 
isotopic shifts. These are just the
lines \ion{C}{i}$^\ast$ $\lambda 1560.68$ 
and $\lambda 1657.90$ mentioned above.
Other \ion{C}{i}$^\ast$ lines
($\lambda 1656.26$ and $\lambda1657.37$), 
as well as the lines with $J=0$ and 2
are less sensitive to the presence of $^{13}$C since their isotopic shifts are
smaller than those of $\lambda 1560.68$ and $\lambda 1657.90$.

Following QBR, we use two-component model to
constrain the isotope ratio ${\cal R}$ from the minimization of the
objective function defined by eq.(2) in Paper~I. 
The $\chi^2_{\rm min}$ values (normalized per degree of freedom)
are calculated in the interval
$30 \leq {\cal R} \leq 3000$ and shown as a function of ${\cal R}$ in Fig.~2. 
This function gradually decreases with increasing ${\cal R}$ and tends
to a limit $\chi^2_{\rm lim} = 1.250$ at ${\cal R} \gg 1$.
This global minimum  $\chi^2_{\rm lim} = 1.250$ lies within
$1\sigma$ uncertainty range since at $\nu = 26$ the expected mean
value of $\chi^2_\nu$ is equal to $1\pm0.277$.
The optimized values and formal uncertainties of the model parameters
for both \ion{C}{i}$^\ast$ absorption components 
(labeled by subscripts `1' and `2') 
at zero $^{13}$C abundance
are as follows:
the column density $N_1 = (3.5\pm0.3)\times10^{13}$ \cm, 
$N_2 = (6.3\pm0.4)\times10^{12}$ \cm,
the Doppler parameter $b_1 = 1.10\pm0.06$ \kms, 
$b_2 = 3.40\pm0.17$ \kms, 
and the radial velocity difference $\Delta v_{2-1} = 9.23\pm0.28$ \kms.
The corresponding synthetic profiles are shown in Fig.~1 by smooth curves.
The model parameters do not change for ${\cal R} > 300$.

The dashed horizontal lines in Fig.~2 mark the $1\sigma$ and $2\sigma$
confidence levels. Since the curve $\chi^2_{\rm min}({\cal R})$ is almost
flat at ${\cal R} > 300$,
the present data
allow us to constrain only the lower limits on the carbon isotopic ratio:
${\cal R} > 235$ ($1\sigma$ C.L.), or a more conservative restriction
${\cal R} > 80$ ($2\sigma$ C.L.). The former value is, however, less
certain due to a small gradient of the $\chi^2_{\rm min}({\cal R})$
function at large ${\cal R}$.

We also used a 3-component model
to test robustness of the obtained results. Being applied to the lines of
\ion{C}{i} $\lambda\lambda1560.3,1658.9$, \ion{C}{i}$^\ast$ 
$\lambda\lambda1560.7,
1657.9$, and \ion{C}{i}$^{\ast\ast}$ 
$\lambda\lambda1561.3, 1657.0, 1658.1$ this
model yields ${\cal R}_{1\sigma} > 145$ and ${\cal R}_{2\sigma} > 78$ with
$\chi^2_{\rm min} = 1.020\,\, (\nu=93)$ at ${\cal R} \gg 1$.
The concordance of all available \ion{C}{i} profiles is 
demonstrated in Fig.~3.

\section{Discussion and Conclusions}

We re-analyzed the profiles of the \ion{C}{i} lines
associated with the sub-DLA system observed at \zabs = 1.15 in the
spectrum of \object{HE 0515--4414}. 
Our main purpose was to set a constraint 
to the carbon isotopic abundance outside the Milky Way, in distant
intervening clouds at the cosmological
epoch corresponding to 8.2 Gyr of the look-back time.
Similar tasks were discussed by Carlsson et al. (1995), 
Labazan et al. (2005), and FMG.

The present $\Delta \chi^2$ analysis gives \iso~$> 80$ (2$\sigma$ C.L.). 
This low abundance of $^{13}$C does not support the enrichment of gas
by the wind from rotating massive stars (Meynet et al. 2005).
Besides, it can constraint the fraction of
intermediate-mass (IM) stars ($4 \la M/M_\odot \la 8$) 
in the IMF.
For instance, the FMG models for the chemical evolution of the
Galaxy with normal IMF attain the solar ratio \iso$ \simeq 90$ already at 
[Fe/H] $\approx -2$ for the outer radius and [Fe/H]
$\approx -1.5$ for the solar radius case, respectively. 
On the contrary,
the models with an enhanced population of IM
stars produce oversolar abundances of $^{13}$C at all
metallicities [Fe/H] $< -0.3$ (cf. Fig.~9 in FMG). 
Thus, these models can be ruled out by our measurements.

Closely related to \iso ratio is the 
problem of nitrogen, $^{14}$N,
which is expected to be produced
in the same way and from about the same stars as $^{13}$C.
Measurements in DLAs show very low N abundances
which is consistent with the absence of 
an enhanced population of
IM stars to the chemical evolution of the gas
(Centuri\'on et al. 2003). Thus, low $^{13}$C and low $^{14}$N
seem to be in agreement. It should be noted, however,
that the chemical evolution of nitrogen is 
still not well understood since the existing models predict
abundances higher than observed in DLAs
(see, e.g., Fig.~6 in FMG). 

The present bound to the abundance of $^{13}$C 
and, as a result, to the contribution of the AGB stars to the chemical
evolution of the \zabs = 1.15 system may be indirectly related to the
claims of a possible time variation of the fine-structure constant, $\alpha$\,
(see Murphy et al. 2004 and references therein). 
A significant portion of the sample used by
Murphy et al. involves the comparison of \ion{Mg}{ii} and \ion{Fe}{ii}
wavelength shifts. Later on, it was shown that 
over-solar abundances ($\sim 0.3$ dex) of 
$^{25,26}$Mg isotopes with respect to $^{24}$Mg
in the absorbing material can imitate an apparent 
variation of $\alpha$ in the redshift range between
0.5 and 1.8 (Ashenfelter et al. 2004a,b).
The production of Mg isotopes is believed to occur in about the same stars
which produce $^{13}$C. Thus, the obtained bound to the amount of $^{13}$C,
taken as a typical value for the DLA systems\footnote{The DLA
population shows a remarkable similarity in the relative chemical
abundances (Prochaska 2003).},
poses a limit to the role of the AGB stars in mimicking the $\alpha$
variations.
However, the variability of $\alpha$ has not been supported by 
more recent studies (Quast et al. 2004; Chand et al. 2004; Levshakov et al.
2005a,b). 
This may suggest the presence of 
other systematic errors to explain the different results 
concerning the variability of $\alpha$. 

\begin{figure*}[t]
\vspace{0.0cm}
\hspace{0.0cm}\psfig{figure=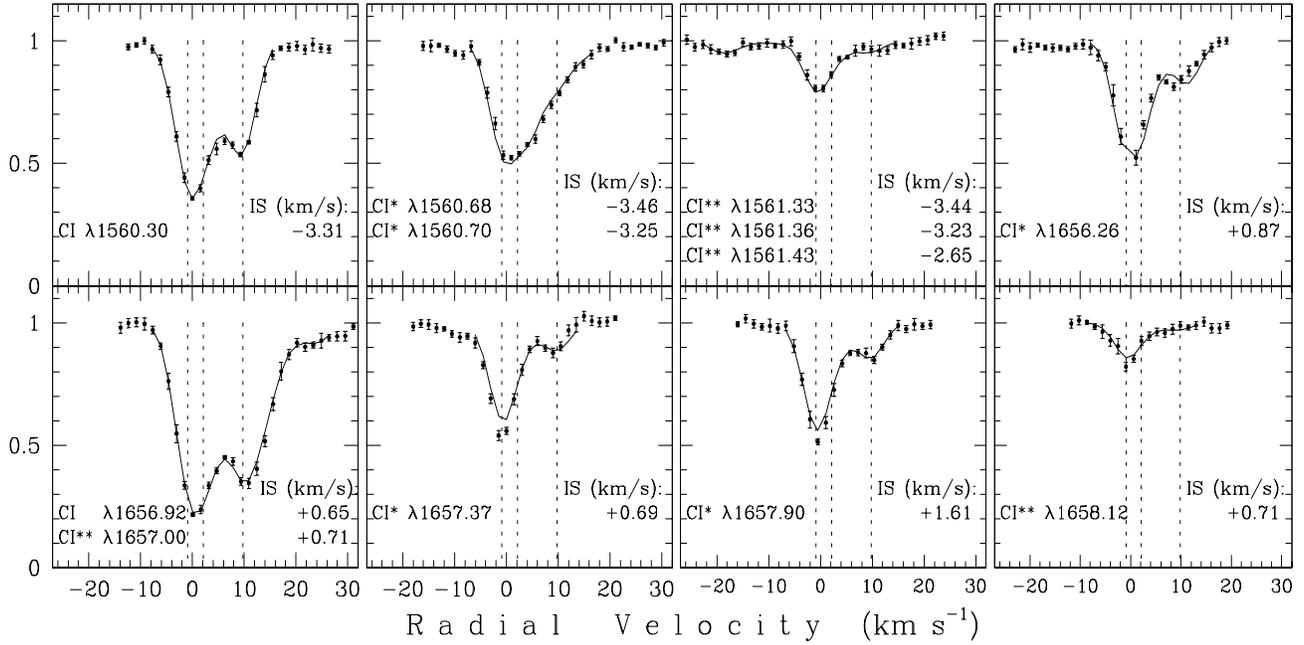,height=10.0cm,width=18.5cm}
\vspace{-0.2cm}
\caption[]{
Same as Fig.~1 but for a 3-component Voigt profile model
fitted to all available \ion{C}{i} lines.
The over-plotted smooth curves show the best
fitted synthetic profiles ($\chi^2_{\rm min} = 1.036$, the number
of degrees of freedom $\nu = 125$). 
The dashed vertical lines mark positions of the \ion{C}{i}
sub-components
}
\label{fig3}
\end{figure*}

\begin{acknowledgements}
The authors thank Edward Jenkins and Ralf Quast for useful comments.
This work is supported by
the RFBR Grant No. 03-02-17522 and by the RLSS Grant No. 1115.2003.2.
\end{acknowledgements}

\end{document}